\documentclass[letterpaper,twocolumn,showpacs,prl,superscriptaddress]{revtex4}
\usepackage{graphicx,color}
\usepackage{amssymb}
\usepackage{amsmath,amsfonts,latexsym}
\usepackage{array,tabularx}
\usepackage{dcolumn}               

\newcommand{\vect}[1]{{\mathbf #1}}
\newcommand{\Frac}[2]{\displaystyle\frac{#1}{#2}}

\begin{document}

\title{Non-equilibrium quantum condensation in an incoherently pumped
  dissipative system}  
\author{M.~H.~Szyma{\'n}ska}
\affiliation{Clarendon Laboratory, Department of Physics, University of Oxford,
             Parks Road, Oxford, OX1 3PU, UK}
\author{J.~Keeling}
\affiliation{Department of Physics, Massachusetts Institute of
  Technology, 77 Mass. Ave., Cambridge, MA 02139, USA}
\author{P.~B.~Littlewood}
\affiliation{Cavendish Laboratory, University of Cambridge,
             Madingley Road, Cambridge CB3 0HE, UK}
\begin{abstract}
  We study spontaneous quantum coherence in an out of equilibrium
  system, coupled to multiple baths describing pumping and decay.  For
  a range of parameters describing coupling to, and occupation of the
  baths, a stable steady-state condensed solution exists.  The
  presence of pumping and decay significantly modifies the spectra of
  phase fluctuations, leading to correlation functions that differ
  both from an isolated condensate and from a laser.
\end{abstract}
\pacs{05.70.Ln, 03.75.Gg, 03.75.Kk, 42.50.Fx}
\maketitle

The phenomenon of condensation, i.e macroscopic occupation of a single
quantum mode, has attracted considerable attention in recent years.
It ranges from Bose-Einstein Condensation (BEC) of structureless
bosons to the BCS-type collective state of fermions and has been
studied in several physical systems such as degenerate atomic gases
and superconductors~\cite{SnokeBook}.
Further, recent experimental advances in manipulation of atomic Fermi
gases have led to realisation of the BCS-BEC crossover regime
\cite{atomicfermi}.
The next challenge is to control and study condensed states
in solid-state.
For this, the currently promising candidates are excitons in coupled
quantum wells~\cite{excitonexp},
microcavity polaritons~\cite{lesidang,deng,richard}, quantum
Hall bilayers~\cite{QH}, and
Josephson junction arrays in microwave cavities~\cite{JJ}.

Unlike atomic gases, solid-state systems face dephasing and decay, as
(with the special exception of superconductors) it
is not usually possible to isolate the condensate from the
environment:
Phonons and impurities lead to dephasing, and due to imperfect
trapping, particles escape, requiring external pumping to sustain a
steady-state.
If such processes are faster than thermalisation the system
remains out of thermal equilibrium.
Dissipation and decay not only present experimental
obstacles, but also pose fundamental questions about
the robustness of a condensate:
Is a steady-state condensate possible with incoherent pumping and
decay, if so, how does it differ from thermal equilibrium? 
Condensation in dissipative systems also provides a connection to the
laser~\cite{Haken}.
The relation between lasing and BEC is particularly relevant for
polariton BEC, where the experimental distinction between the two is
not straightforward~\cite{marz}.

Models that combine potentially strong
non-equilibrium pumping with spontaneous symmetry breaking are not
well-studied, which motivates a study of simple models to extract the
principal qualitative features.  In this Letter we study spontaneous
condensation in a system coupled to independent baths, not in thermal
or chemical equilibria with each other, providing incoherent pumping
and decay.
We focus on a Bose-Fermi system with disorder localised fermions; this
is a model for exciton-polaritons~\cite{lesidang,deng,richard} or
Josephson junctions in microcavities~\cite{JJ}.
However, many of our conclusions apply more generally to condensation
with pumping and decay.
We show that steady-state spontaneous condensation can occur in such
systems, and can be distinct from lasing:
The condensate can exist at low densities, far from the inversion
required for lasing.
We study fluctuations about a steady-state condensate and find that
the collective modes are qualitatively altered by the presence of
pumping and decay:
The low energy phase mode (Goldstone, Bogoliubov mode) becomes
diffusive at small momenta.
By considering the effect of phase fluctuations, we find the decay of
correlations, which at large times and distances 
differs both from that for a thermal equilibrium
condensate and from a laser.

Our Hamiltonian is
$\hat{H} = \hat{H}_{sys}+\hat{H}_{sys,bath}+\hat{H}_{bath},$ 
where,
\begin{multline}
  \hat{H}_{sys} 
  = 
  \sum_{\alpha} \varepsilon_{\alpha}
  \left(
    b_{\alpha}^\dag b_{\alpha}^{} - a_{\alpha}^{\dag}a_{\alpha}^{}
  \right)
  + 
  \sum_{\vect{p}} 
  \omega^{}_{\vect{p}} \psi_{\vect{p}}^\dag \psi_{\vect{p}}^{}
\\ 
  + 
  \Frac{1}{\sqrt{L^2}} 
  \sum_{\alpha} \sum_{\vect{p}} \left(
    g^{}_{\alpha, \vect{p}} \psi_{\vect{p}}^{} b_{\alpha}^\dag a_{\alpha}^{} +
    \text{h.c.}
  \right), 
\label{Hsys}
\end{multline}
describes two fermionic species $b_{\alpha}$ and $a_{\alpha}$,
interacting with bosonic modes $\psi_{\vect{p}}$, and coupled to
three baths;
\begin{multline} 
  \hat{H}_{sys,bath}
  = 
  \sum_{\alpha,k}
  \Gamma^{a}_{\alpha,k} (a^\dagger_\alpha A_{k}^{} +
  \text{h.c.} ) +
  \Gamma^{b}_{\alpha,k}
  (b^\dagger_\alpha B_{k}^{} + \text{h.c.}) 
  \\ 
  +
  \sum_{\vect{p},k}
  \zeta^{}_{\vect{p},k}(\psi_{\vect{p}}^\dagger \Psi_{k}^{} +
  \text{h.c.}),
\label{Hsysbath}
\end{multline} 
given by
$\hat{H}_{bath}  = 
\sum_{k}\omega^{\Gamma^a}_{k}
A_{k}^{\dagger}A_{k}^{} +
\sum_{k}\omega^{\Gamma^b}_{k}
B_{k}^{\dagger}B_{k}^{} +
\sum_{k}\omega^\zeta_{k}\Psi_{k}^\dagger \Psi_{k}^{}.$
%
A single two level system coupled to multiple baths has been recently
studied in the context of the Kondo problem \cite{Paaske}. Condensed
solutions of eq.~(\ref{Hsys}) have been studied in the context of
atomic Fermi gases \cite{Holland} and microcavity
polaritons~\cite{jonathan,fran}.
In this Letter we focus on microcavity polaritons, so
$b^{\dagger}_{\alpha}, a^{}_{\alpha}$ describe an electron and hole
within a disorder-localised exciton state of energy
$\varepsilon_{\alpha}$. This can also be viewed as a
fermionic representation of a hard-core boson or of a spin.
These are dipole coupled to cavity photon modes $\psi_{\vect{p}}$, with
low $\vect{p}$ dispersion, $\omega_{\vect{p}} \simeq \omega_0 +
\vect{p^2}/2 m_{\text{ph}}$, where $m_{\text{ph}} = (\hbar/c)(2\pi/w)$
is the photon mass in a 2D microcavity of width $w$.
Due to the finite reflectivity of the cavity mirrors, photons escape,
so the system must be pumped (excitons injected) to sustain a
steady-state.
Incoherent fermionic pumping and photon decay are described by
(\ref{Hsysbath}) where $A_{k}^{}, B_{k}^{}$ are fermionic annihilation
operators for the pump baths, while $\Psi_{k}^{}$ are bosonic
annihilation operators for photon modes outside the cavity.

With pumping and decay, i.e. coupling to baths with different
temperatures and/or chemical potentials, the system distribution
function can be far from thermal and needs to be obtained
self-consistently with the system's spectrum.
Therefore we use non-equilibrium field theory in the Keldysh
path-integral formulation~\cite{kamenev}.
Integrating over the bath's degrees of freedom and fermion fields
yields an effective action for photons.
The dependence of the effective action on the bath properties is
parameterised by the functions $\kappa(\omega)$, $\gamma(\omega)$,
$F_{A,B}(\omega)$, and $F_\Psi(\omega)$.
The cavity decay rate
$\kappa(\omega)=\pi\zeta^2(\omega)N^{\zeta}(\omega)$, where $\zeta$ is
the coupling of the cavity photons to the bosonic modes in
eq.~(\ref{Hsysbath}) and $N^{\zeta}$ is the density of states of these
modes.
Similarly $\gamma(\omega)=\pi \Gamma^2(\omega) N^\Gamma(\omega)$ where
$\Gamma$ and $N^\Gamma$ are respectively coupling to, and the density
of states of, the fermionic pumping baths.
In this work we assume a flat spectrum for the baths, so $\kappa$ and
$\gamma$ are frequency independent.
Frequency dependence is however present in the bath distribution
functions; $F_{A,B}(\omega)=1-2n^{A,B}(\omega)$, and
$F_{\Psi}(\omega)=1+2n^{\Psi}(\omega)$, where $n^{A,B},n^{\Psi}$
are occupations of the baths.

We proceed by extremising the non-equilibrium effective action with a
steady-state, uniform, photon field of the form,
\begin{math}
  \psi(t)=\psi e^{-i\mu_S t},
\end{math}
giving a non-equilibrium generalisation of the usual gap equation (e.g.
\cite{jonathan,nagaosa}):
\begin{equation}
  (\omega_{0}-\mu_S-i\kappa)\psi =g \mathrm{Tr} (iG_{ba}^{K}).
  \label{Gap}
\end{equation}
Here,
\begin{math}
  iG_{ba}^{K}(t,t')
  =
  \langle a_{cl}^{\dagger}(t)b_{cl}^{}(t') \rangle,
\end{math}
is the anomalous Keldysh fermionic Green's function, where
$a_{cl}=(a_f+a_b)/\sqrt{2}$ and $f$, $b$ are the forward and backward
branches of the Keldysh time contour \cite{kamenev}, and is given by:
\begin{equation}
  iG^K_{ba}(\omega)
  = 
  2\gamma g \psi 
  \frac{(F_A+F_B)\omega+(F_B-F_A)(\tilde{\epsilon}+i\gamma)}
  {[(\omega-E)^2+\gamma^2][(\omega+E)^2+\gamma^2]},
  \label{GK}
\end{equation}
where
$E=\sqrt{\tilde{\epsilon}^2+g^2|\psi|^2}$, $\tilde{\epsilon}=\epsilon -
\mu_S/2$, and the arguments of $F_{B,A}(\omega)$ are
shifted by $\pm \mu_S/2$ while $\omega$ is a real frequency.
The bosonic bath's distribution $F_{\Psi}$ does not enter the
mean-field gap equation, as the mean-field considers only the
condensed photons, and not the population of incoherent photons.
In the limit $\gamma, \kappa \to 0$ with the bath distributions
$F_{A,B}$ being thermal, eq.~(\ref{Gap}) reduces to its equilibrium
form~\cite{jonathan,fran}; for finite $\gamma$ and $\kappa$ it
is significantly altered.

As in thermal equilibrium, the normal state $\psi=0$ is always a
solution of eq.~(\ref{Gap}), but for some range of parameters there is
also a condensed $\psi \ne 0$ solution.
When a solution $\psi \ne 0$ exists, the solution $\psi=0$ becomes
unstable.
To understand this instability, we consider small fluctuations about
the mean-field, $\psi=\psi_0+\delta \psi$.
The effective action for these fluctuations $\delta \psi$ has a part
from the free photon action, and a part from interactions with
fermions, $\frac{1}{2}\mathrm{Tr}(G\delta G^{-1} G \delta G^{-1})$, where
the fermionic Green's functions $G$ are four by four matrices in the
Keldysh and particle-hole ($a$, $b$) spaces.
Inverting the effective action for fluctuations gives the photon
Green's functions,
\begin{math} 
  \mathcal{D}^{K,R,A}.
\end{math}
The retarded and advanced Green's functions $\mathcal{D}^{R,A}$ give
the excitation spectrum.
The distribution function
\begin{math}
  {F}_S
\end{math}
, defined by
\begin{math}
  \mathcal{D}^{K}
  =
  \mathcal{D}^{R}{F}_S-{F}_S\mathcal{D}^{A},
\end{math}
determines how the spectrum is occupied.
As the system need not be in thermal or chemical equilibria with the
baths, ${F}_S$ is in general not thermal, and differs from the
bath distributions, $F_{A,B}$ and $F_{\Psi}$.

The instability of the normal state when a condensed solutions exists
is analogous to that in thermal equilibrium.
Even when the bosonic distribution ${F}_S$ is far from thermal, as the
system approaches the phase transition, ${F}_S$ will have a divergence
at a frequency which we define as an effective chemical potential,
where $\Im({\mathcal{D}^{R}}^{-1}(\mu_{\mbox{eff}}))=0$.
Taking the zeros of $\Re({\mathcal{D}^{R}}^{ -1}(\omega^{\ast},q))$ as
defining the normal modes of the system, condensation occurs when
$\mu_{\mbox{eff}}$ reaches the bottom of the band of excitations, as
in equilibrium.
This condition, that a solution to
${\mathcal{D}^{R}}^{-1}(\mu_{\mbox{eff}},q=0)=0$ exists, is equivalent
to the gap equation, eq.~(\ref{Gap}), at $\psi=0$, $\mu_{\mbox{eff}} =
\mu_{S}$, since ${D}^{R}$ describes a susceptibility which diverges at
the transition.
Beyond this point, the normal state is unstable, as $\mu_{\mbox{eff}}$
would lie in a bosonic band.
This idea of instability can be directly connected to another
definition: 
Beyond this point, the poles of the $\mathcal{D}^R$ have positive
imaginary parts, fluctuations grow (rather than
decay) in time exponentially.
To see this, consider the imaginary parts of the poles of
$\mathcal{D}^R$ as a function of momentum, $q$.
At large $q$, these poles describe bare photons, and so are stable.
By the previous definitions, above the transition, there is a
$q$ at which the normal state Green's function has a real pole, 
${\mathcal{D}^{R}}^{-1}(\omega^{\ast},q)=0$.
The sign of the imaginary part of the pole changes at this point, so
the low $q$ poles are unstable.
Thus, whenever there is a condensed solution to the gap equation the
$\psi = 0$ solution is always unstable.

There is, however, an important difference between the non-equilibrium
steady-state and thermal equilibrium.
Unlike thermal equilibrium, there is a range of parameters for which
neither the normal state nor condensed solutions of the form
\begin{math}
  \psi(t)=\psi e^{-i\mu_S t},
\end{math}
are stable.
This is not too surprising as  systems similar to
(\ref{Hsys}) are known to follow a complicated or even chaotic
dynamics~\cite{chaos}.
Although such anharmonic solutions would be of a great interest, we
focus here on steady-state condensed solutions of the usual form.

Having understood the stability of solutions, we next solve the gap
equation~(\ref{Gap}) to find the non-equilibrium phase diagram.
Since eq.~(\ref{Gap}) is complex it gives two equations for two
unknowns: the order parameter $\psi$ and the frequency $\mu_S$.
The common oscillation frequency $\mu_S$ would in thermal equilibrium
be the system's chemical potential, considered as a control parameter,
adjusted to match the required density, and the (real) gap equation
determines only $\psi$.
Here, because different baths have different chemical potentials, the
system is not in chemical equilibrium with either bath, so both
$\mu_S$ and $\psi$ must be found from the gap equation.
The density, which can be found given $\psi$ and $\mu_S$,
is set by the relative strength of the pump and decay.

From here, the bath distributions are taken to be individually in thermal
equilibrium, although the formalism would allow any distribution.
However, as the baths need not be in equilibrium with each other, the
system can still be far from thermal equilibrium.
Further, for simplicity, the figures shown are all for the baths at
zero temperature, so the bath distributions are defined by their
chemical potentials.
To ensure that on average only one of the two fermionic levels $a,b$
is occupied, the chemical potentials of baths $A,B$ are related by
$\mu_A = - \mu_B$.
Since the cavity photon modes start at energies much above the zero
for bulk photon modes, we take the chemical potential of the decay
bath to be large and negative.
With these restrictions, the baths are described by three parameters,
the couplings $\gamma, \kappa$ and $\mu_B$ parameterising the
occupation of the pumping baths.

\begin{figure}
  \begin{center}
    \includegraphics[width=1\linewidth,angle=0,clip]{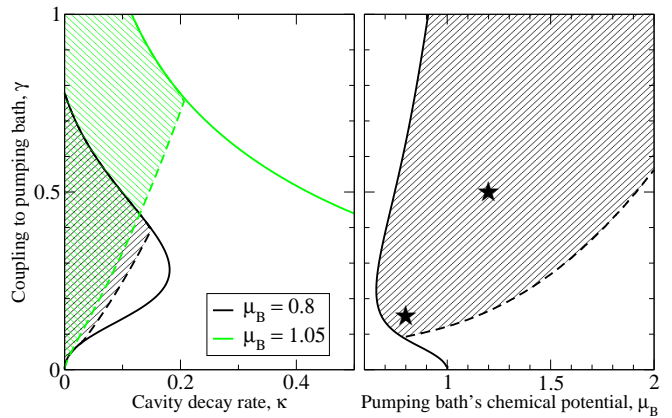}
  \end{center}
  \caption{ Mean field phase diagrams.  Left: Critical $\gamma$ vs
    decay rate $\kappa$, for two different pumping bath chemical
    potentials, $\mu_B$. Right: Critical $\gamma$ vs $\mu_B$ for
    $\kappa$=0.038g as in experiment \cite{richard}. Solid lines
    show where a solution of the gap equation exists, dashed lines
    show where that solution is stable.  The region where a stable
    condensed solution exists is shaded.  Stars mark the choice of
    parameters shown in Fig.~\ref{fig:lum}.}
  \label{fig:phase}
\end{figure}

Figure~\ref{fig:phase} shows a phase diagram in terms of the
parameters $\gamma, \kappa, \mu_B$.
It shows both where a condensed solution to the gap equation exists,
and where the solution is stable.
At a given decay rate $\kappa$ there is a minimum $\gamma$ (as pumping
is proportional to $\gamma$) and a maximum $\gamma$ (as dephasing is
also proportional to $\gamma$) required for condensation.
Stable condensed solutions exist only for $\kappa$ smaller than about
$0.2g$.
In the region shown in Fig.~\ref{fig:phase} the condensed solutions
are all below population inversion.
However for $\gamma>g$, when system is in a weak coupling regime, only
laser-like solutions which require population inversion are possible.
For large $\gamma$ and $\mu_B$ (large pumping), our
  theory recovers the regular laser limit.

We now discuss the collective modes, considering fluctuations about
the steady-state.
Motivated by microcavity polaritons, we study these collective modes
by calculating the photoluminescence (PL) spectra:
\begin{math}
  i\mathcal{D}_{\psi^{\dagger}\psi^{}}^{<}(t,t')
  =
  \langle\psi_f^{\dagger}(t)\psi_b^{}(t')\rangle
\end{math}.
In the normal state, as expected, one finds homogeneously broadened
upper and lower polariton modes.
This broadening depends on all of the bath parameters, $\kappa, \gamma$
and $\mu_B$.
Approaching the phase boundary from the normal side, the lower
polariton linewidth reduces to zero.
This can be understood by identifying the zeros of the real part of
the inverse Green's function as the polariton energies,
\begin{math}
  \Re({\mathcal{D}^{R}}^{ -1}(\omega^{\ast},q)) = 0,
\end{math}
and the imaginary part as giving the linewidth,
\begin{math}
  1/\tau_p \approx \Im({\mathcal{D}^{R}}^{-1}(\omega^{\ast},q)).
\end{math}
The earlier discussion relating the gap equation to
${\mathcal{D}^{R}}^{ -1}(\mu_S,q=0)$ implies that $1/\tau_p$
vanishes at the transition.


In the condensed state, as in equilibrium\cite{jonathan}, the
excitation spectrum changes.
This leads to a soft mode, describing phase fluctuations, as
global phase rotation symmetry is broken.
These phase fluctuations may be large, so one must include the phase
fluctuations to all orders~\cite{nagaosa} to find the
field-field correlations; amplitude fluctuations remain small, as they
have a restoring force.
Writing,
\begin{math}
    \psi(t) = \sqrt{\rho_0 + \pi(t)} e^{i\phi(t)}, 
\end{math} 
where $\rho_0$ is the mean-field condensate density, the PL spectrum
is thus:
\begin{eqnarray}
  \label{eq:field-correlations}
  i \mathcal{D}^{<}_{\psi^{\dagger}\psi}(t,r)
  = 
  \rho_0 \left[
    1
    +
    \mathcal{O}\left(1/\rho_0\right)
  \right]
  \exp\left[ -f(t,r) \right],
  \\
  \label{eq:correlator-exponent}
  f(t,r)
  =
  \int d\nu  
  \int(dq)^2
  \left[
    1-
    e^{ i(\nu t + \vect{q}\cdot\vect{r}) }
  \right]
  i \mathcal{D}^{<}_{\phi\phi}(\nu,q).
\end{eqnarray}
The phase-phase Green's function $\mathcal{D}^{<}_{\phi\phi}$ is found
by inverting the action expanded to quadratic order.

\begin{figure}
  \begin{center}
    \includegraphics[width=1\linewidth,angle=0,clip]{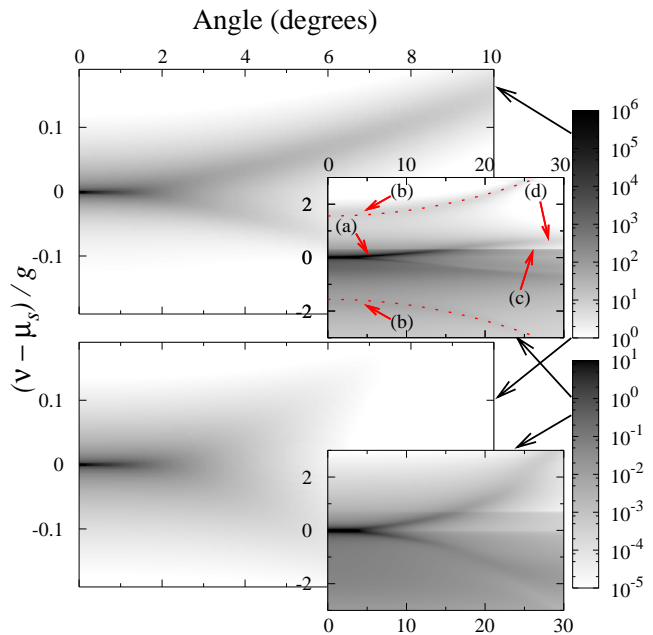}
  \end{center}
  \caption{Photoluminescence $i
    \mathcal{D}^{<}_{\psi^{\dagger}\psi}(\nu,q)$ of a condensed
    system, where $q$ is shown by angle of emission
    $\rm{tan}^{-1}(cq/\omega_0)$. Top: Strong coupling, Bottom: weak
    coupling, exact parameters marked by stars on
    Fig.~\ref{fig:phase}.  Main figures show PL from small $\nu,q$
    region, to all orders in phase fluctuations.  Insets show a larger
    range of $\nu,q$ for the same parameters.  (Dotted lines have been
    added to show the faint amplitude mode).  }
\label{fig:lum}
\end{figure}

The $\mathcal{O}\left(1/\rho_0\right)$ terms in
eq.~(\ref{eq:correlator-exponent}), due to amplitude-amplitude and
phase-amplitude Green's functions, are included when plotting
Fig.~\ref{fig:lum}.
The form of these terms, and the full expression for
$\mathcal{D}^{<}_{\phi\phi}$, are  complex, so we do not
reproduce them here.
The behaviour of $\mathcal{D}^{<}_{\phi\phi}$ in the limit of small
frequencies and momenta, is of interest, and takes a simple
form:
\begin{equation}
  \label{eq:phase-greens-function}
  i \mathcal{D}^{<}_{\phi\phi}(\nu,q)
  \approx
  C \left[
     \left( c^2 q^2 -  \nu^2\right)^2 + 4 \nu^2 x^2
   \right]^{-1},
\end{equation}
where $C$, $c$, and $x$ can be found from the full expressions.
Without pumping and decay, the term $x$ would vanish.
Its presence means that rather than a linear dispersion at low $q$, as
in a closed system in thermal equilibrium\cite{nagaosa,jonathan}, the
poles of the Green's function have the following form $\nu=-ix \pm
i\sqrt{x^2-c^2q^2}$.
For $|q|<x/c$, these modes are diffusive; the poles are entirely
imaginary, only for $|q|>x/c$, do they acquire a real part.
This behaviour is apparent in the PL spectrum in Fig.~\ref{fig:lum}.
The diffusive behaviour at small $q$ is not limited to Bose-Fermi
systems; it should also be present in other condensed systems with
pumping and decay.
Note that at $q=0$ and $\nu=0$, $\mathcal{D}^{<}_{\phi\phi}$ has a
real pole; a manifestation of broken symmetry in the infinite system.

As given by eq.~(\ref{eq:field-correlations}), the PL spectrum does not
depend linearly on $\mathcal{D}^{<}_{\phi\phi}$.
If $\mathcal{D}^<_{\phi\phi}$ were small, the exponential in
eq.~(\ref{eq:field-correlations}) could be expanded, giving PL divided
between a condensate term $\rho_0\delta(\nu)\delta(q)$, and a part
from occupation of the fluctuation spectrum.
However, at $\nu, q \to 0$, $\mathcal{D}_{\phi\phi}$ is not small, and
so phase fluctuations give a lineshape to the condensate and determine
long time field-field correlations.
In 2D, inserting eq.~(\ref{eq:phase-greens-function}) in
eq.~(\ref{eq:correlator-exponent}), the main dependence of $f(t,r)$ on
$t,r$ comes from a logarithmic divergence, $\int dq /q$, cut at large
momenta by terms beyond those in eq.~(\ref{eq:phase-greens-function}),
and at small momenta by one of $1/r, 1/ct, 1/c\sqrt{t\tau_g}$ or
$1/c\tau_g$, where $\tau_g=1/x$ is the lifetime of the
phase mode.
This logarithmic form leads to power law field-field correlations.
(This is true also in thermal-equilibrium\cite{nagaosa}, for
which,
\begin{math}
  f(t,r)= \eta \ln \left( \sqrt{c^2 t^2 + r^2} / \beta c \right),
\end{math}
with $\eta \propto k_B T / \rho_0$.)
According to the relative values of $r, t, \tau_g$, different lower
cutoffs apply, and so the power law differs at different places in the
$r,t$ plane.
For small $r,t$ we recover the equilibrium power laws; however when $
t \gg \tau_g$, and $ c \sqrt{t \tau_g} \gg r$, then $\tau_g$ becomes
important, and
\begin{math}
  f(t) \propto  \ln( c \sqrt{t \tau_g}),
\end{math}
giving a condensate lineshape that differs from that of an
non-dissipative, thermal equilibrium 2D system.
Power law field correlations at long times lead to a power law
divergence of their Fourier transform as $\nu,q \to 0$, so there is no
well defined linewidth.
This differs from phase diffusion for a single mode studied in Laser
theory\cite{Haken}, where phase correlations grow linearly in time,
giving exponential decay of field correlations.


At higher $q$ and $\nu$ (Fig \ref{fig:lum} Insets) the difference
between expanding to linear order in Green's functions and keeping all
orders of phase fluctuations is not visible.
The strong coupling spectrum consists of the usual phase and amplitude
modes (marked a,b) emerging from the lower and upper polariton
branches.
One can also see an occupation edge (c), and above that the lower
polariton following the exciton dispersion (d).
These features at higher momenta are analogous to thermal equilibrium
non-dissipative polariton condensation in strong coupling
regime~\cite{jonathan} but with non-thermal occupations.
Increasing pumping, dephasing increases, and the system crosses to
weak-coupling (lower panels of Fig.~\ref{fig:lum}), the polariton
splitting is suppressed, and the spectrum follows the photon
dispersion.
The diffusive region, then linear dispersion, at small momentum, seen
in the main figures, occurs both in strong and weak coupling and is a
sign of condensation in the dissipative system.


To conclude, we have studied how steady-state spontaneous condensation
emerges in non-equilibrium systems with pumping and decay.
This condensation is distinct from lasing:
It can occur at densities much lower than the population inversion
required for lasing, and the decay of correlations, and thus
condensate lineshape, differ from that for phase diffusion of a single
laser mode. 
Dissipation qualitatively changes the
fluctuation spectrum with respect to isolated thermal
equilibrium:
At low momenta, the phase mode becomes diffusive, changing the
power-laws controlling long time decay of field-field correlations.
These conclusions, although studied here for microcavity polaritons,
apply also to other Bose and Bose-Fermi condensates subject to pumping
and decay.

\begin{acknowledgments}
We are grateful to B. D. Simons and R. Zimmermann for 
stimulating discussions. We acknowledge financial support from 
EPSRC (MHS) and the Lindemann Trust (J.K).

\end{acknowledgments}


\begin{thebibliography}{99}
%
\bibitem{SnokeBook} 
  A. Griffin, D. W. Snoke, and S. Stringari, eds.,
  \emph{Bose-Einstein Condensation},
  (Cambridge University Press, Cambridge, 1995).

\bibitem{atomicfermi}
  C. A. Regal, M. Greiner, and D. S. Jin, 
  Phys. Rev. Lett. \textbf{92}, 040403 (2004);
  %
  M. W. Zwierlein \emph{et al.},
  Phys. Rev. Lett. \textbf{92}, 120403 (2004).

\bibitem{excitonexp}
  L. V. Butov \emph{et al.}, 
  {\em Nature}, {\bf 417}, 47 (2002); {\bf 418}, 751 (2002);
  D. Snoke \emph{et al.}, 
  {\em Nature}, {\bf 418}, 754  (2002).

\bibitem{lesidang}
  %
  Le Si Dang \emph{et al.}, Phys. Rev. Lett. \textbf{81}, 3920 (1998).

\bibitem{deng} 
  %
  H. Deng \emph{et al.}, Science \textbf{298}, 199 (2002);
  PNAS \textbf{100}, 15318 (2003).

\bibitem{richard}
  %
  M. Richard \emph{et al.}, Phys. Rev. Lett. \textbf{94}, 187401 (2005).

\bibitem{QH} 
  J. P. Eisenstein, A. H. MacDonald {\em Nature} \textbf{432}, 691
  (2004).

\bibitem{JJ}
  P. Barbara \emph{et al.},
  Phys. Rev. Lett. \textbf{82}, 1963 (1999).

\bibitem{Haken}
  H. Haken, in \emph{Quantum Optics}, edited by Kay and Maitland
  (academic Press, London, 1970), p. 201.

\bibitem{marz} 
  M. H. Szymanska and P.B. Littlewood,
  Sol. Stat. Comm. \textbf{124}, 103 (2002); 
  M. H. Szymanska, P.B. Littlewood, and B. D. Simons
  Phys. Rev. A \textbf{68}, 013818 (2003).

\bibitem{Paaske}
 J. Paaske \emph{et al.}, Phys. Rev. B \textbf{70}, 155301 (2004) 

\bibitem{Holland}
  M. Holland, \emph{et al.},
  Phys. Rev. Lett. \textbf{87}, 120406 (2001);
  E. Timmermans, \emph{et al.},
  Phys. Lett. A \textbf{285}, 228 (2001);
  Y. Ohashi, A. Griffin, 
  Phys. Rev. Lett. \textbf{89}, 130402 (2002).

\bibitem{jonathan} 
  J. Keeling, \emph{et al.}, Phys. Rev. Lett. \textbf{93}, 226403 (2004);
  Phys. Rev. B \textbf{72}, 115320 (2005).

\bibitem{fran}
  F. M. Marchetti \emph{et al.}, 
  Phys. Rev. Lett. \textbf{96}, 066405 (2006).
 
\bibitem{kamenev}
  A. Kamenev, in \emph{ Nanophysics: Coherence and Transport}
  (Elsevier, 2005), p 177.

\bibitem{nagaosa}
 N. Nagaosa, \emph{Quantum field theory in strongly correlated
 electronic systems} (Springer-Verlag, 1999).

\bibitem{chaos}
C. Emary, T. Brandes, Phys. Rev. Lett. \textbf{90}, 044101 (2003).



\end{thebibliography}

\end{document}